\documentclass[prd,twocolumn,floatfix,superscriptaddress,preprintnumbers]{revtex4}

\usepackage{graphicx}
\input{epsf}
\usepackage{epsf}
\usepackage{graphicx,epsfig}
\usepackage{bm}
\usepackage{latexsym,amssymb,amsmath,amsfonts,float}

\newcommand {\ga} {\ {\raise-.5ex\hbox{$\buildrel>\over\sim$}}\ }
\newcommand {\la} {\ {\raise-.5ex\hbox{$\buildrel<\over\sim$}}\ }
\newcommand{\eqn}[1] {Eq.~(\ref{#1})}

\def\be{\begin{equation}}
\def\ee{\end{equation}}
\def\ba{\begin{eqnarray}}
\def\ea{\end{eqnarray}}
\renewcommand{\(}{\left(}
\renewcommand{\)}{\right)}
\renewcommand{\[}{\left[}
\renewcommand{\]}{\right]}
\def\ve{\varepsilon}

\begin{document}

\title{Dark energy from a quintessence (phantom) field
rolling near potential minimum (maximum)}

\author{Sourish Dutta}
\email{sourish.d@gmail.com}
\affiliation{Department of Physics and Astronomy, Vanderbilt University,
Nashville, TN  ~~37235}

\author{Emmanuel N. Saridakis }
\email{msaridak@phys.uoa.gr}
\affiliation{Department of Physics,
University of Athens, GR-15771 Athens, Greece}

\author{Robert~J. Scherrer}
\email{robert.scherrer@vanderbilt.edu}
\affiliation{Department of Physics and Astronomy, Vanderbilt University,
Nashville, TN  ~~37235}

\begin{abstract}
We examine dark energy models in which a quintessence or a phantom
field, $\phi$, rolls near the vicinity of a local minimum or
maximum, respectively, of its potential $V\(\phi\)$. Under the
approximation that $(1/V)(dV/d\phi) \ll 1$, [although $(1/V)(d^2
V/d\phi^2)$ can be large], we derive a general expression for
the equation of state parameter $w$
as a function of the scale factor for these models. The dynamics
of the field depends on the value of $\(1/V\)\(d^2 V/d\phi^2\)$
near the extremum, which describes the potential curvature. For
quintessence models, when $\(1/V\)\(d^2 V/d\phi^2\)<3/4$ at the
potential minimum, the equation of state parameter $w(a)$ evolves
monotonically, while for $\(1/V\)\(d^2 V/d\phi^2\)>3/4$, $w(a)$
has oscillatory behavior.  For phantom fields, the dividing line
between these two types of behavior is at $\(1/V\)\(d^2
V/d\phi^2\) = -3/4$. Our analytical expressions agree within 1\%
with the exact (numerically-derived) behavior, for all of the
particular cases examined, for both quintessence and phantom
fields.  We present observational constraints on these models.
\end{abstract}

\maketitle

\section{Introduction}

It has been known for over a decade \cite{Knop,Riess1} that at
least 70\% of the energy density in the universe is in the form of
an exotic, negative-pressure component, called dark energy (see
Ref. \cite{Copeland} for a recent review). The equation of state
of this dark component, defined as the ratio of its pressure to
its density:
 \be
  w=p_{\rm DE}/\rho_{\rm DE},
 \ee is observationally
constrained to be close to $-1$. In particular, constraints
on constant $w$ lead typically to $w=-1$ with $\pm 10\%$ accuracy
\cite{Kowalski:2008ez,Wood-Vasey,Davis}. However, a variety of
cosmological paradigms, based on scalar fields, attribute a
dynamical nature to dark energy, leading to a time-varying  $w$.
The class of models in which the scalar field is canonical is
dubbed quintessence
\cite{RatraPeebles,Wetterich88,TurnerWhite,CaldwellDaveSteinhardt,LiddleScherrer,SteinhardtWangZlatev}
and has been extensively studied in the literature. An alternative
approach is phantom dark energy, a component for which
$w<-1$, which can be realized by using a scalar field with a
negative kinetic term in its Lagrangian, as first proposed by
Caldwell \cite{Caldwell}. Such models have well-known problems
\cite{CarrollHoffmanTrodden,ClineJeonMoore,BuniyHsu,BuniyHsuMurray},
but nevertheless have also been widely studied as potential dark
energy candidates
\cite{Caldwell,Guo,ENO,NO,Hao,Aref,Peri,Sami,Faraoni,Chiba,Setare:2008pc,KSS,Saridakis:2008fy}.
Finally, since both quintessence and phantom models cannot lead to
a dark energy equation-of-state parameter that crosses $-1$ during
cosmological evolution (which might be the possibility according
to observations), the simultaneous consideration of both models in
a joint scenario, named quintom, has recently gained significant
attention
\cite{Feng05,Guo05,Feng06,Cai07,Setare:2008pz,Setare:2008si}.

Given the considerable freedom that exists in choosing the
potential function $V\(\phi\)$ of the scalar field, it is useful
to develop general expressions for the evolution of $w$ which
cover a wide range of models. A general result of this type was
derived in \cite{ScherrerSen1} for a class of quintessence models
in which $w$ is assumed to be always close to $-1$ and the
potential is nearly flat. The flatness of the potential is
characterized by two ``slow roll conditions'':
\begin{equation}
\label{slow1}
\left(\frac{1}{V} \frac{dV}{d\phi}\right)^2 \ll 1,
\end{equation}
and
\begin{equation}
\label{slow2}
\frac{1}{V}\frac{d^2 V}{d\phi^2} \ll 1.
\end{equation}
For these models, it was shown in \cite{ScherrerSen1} that the
evolution of $w$ is described by a unique expression involving
only the present values of $\Omega_{\phi}$ and $w$.  In
\cite{ScherrerSen2} this result was extended to phantom models
satisfying (\ref{slow1})-(\ref{slow2}), and the corresponding $w$
dependence was shown to be described by the same expression with
an overall sign change. Finally, the extension to quintom scenario
under (\ref{slow1})-(\ref{slow2}) was performed in
\cite{Setare:2008sf}, where a universal expression for $w$ was
also extracted, allowing for crossing of $w=-1$.

While the slow-roll conditions (\ref{slow1}) and (\ref{slow2}) are
sufficient to give $w\simeq-1$ today, they are not necessary.
Classes of models characterized by the validity of (\ref{slow1})
alone, i.e., without (\ref{slow2}), were considered in
\cite{DuttaScherrer}, corresponding in particular to a
quintessence field in the vicinity of a local maximum of its
potential. In this case, there is an extra degree of freedom,
namely the value of $(1/V)(d^2 V/d\phi^2)$, which describes
the curvature of the potential in the vicinity of the
maximum. As a result, instead of a single solution for the
evolution of $w$ one obtains a family of solutions that depend on
the present-day values of $\Omega_\phi$ and $w$ and the value of
$(1/V)(d^2 V/d\phi^2)$ at the maximum of the potential. As
expected, this family of solutions includes the slow-roll solution
of \cite{ScherrerSen1} as a special case in the limit where
$(1/V)(d^2 V/d\phi^2) \rightarrow 0$. In \cite{DuttaScherrer1}, a
similar  result was derived for a phantom field evolving near a
minimum of its potential. It was found that a unique family of
solutions, very similar to the one derived in
\cite{DuttaScherrer}, can be used to approximate the behavior of
$w$ in these models.

More recently, Chiba \cite{Takeshi} showed that
while conditions (\ref{slow1})
and (\ref{slow2}) are sufficient to
allow the quintessence version of the slow-roll conditions
to be applied, they are not necessary.  Chiba extended the
methodology of Ref. \cite{DuttaScherrer}
to provide a more general set
of conditions on the potential, dropping the assumption
that the field is close to a local maximum
in the potential, while still assuming
that $w$ is close to
$-1$ throughout the evolution. Interestingly, the
expression derived under these more general conditions coincides
exactly with the expression derived in \cite{DuttaScherrer},
indicating the generality of this result.

In this work we extend the techniques developed in
\cite{DuttaScherrer} to the ``opposite" of the case considered
there.  In Ref. \cite{DuttaScherrer}, the quintessence field was
considered to roll near a local maximum in the potential, so that
$V^{\prime\prime} >0$. In this paper, we consider the case for
which $V^{\prime\prime} < 0$. This corresponds to a canonical
scalar field rolling close to a minimum of its potential.  For
completeness, we also examine the corresponding phantom model, for
which the phantom field rolls near a maximum of its potential.
That is, while in \cite{DuttaScherrer} and \cite{DuttaScherrer1}
the field rolls away from an unstable potential-extremum, in the
present work the field rolls towards a stable extremum. (Recall
that phantom fields roll away from local minima and are attracted
by local maxima). As we will see, this investigation yields a more
complicated dynamics for the scalar field, which can include
oscillations in addition to slow roll behavior.  For simplicity,
we assume that condition (\ref{slow1}) is satisfied, and that the
field evolves near an extremum in the potential, rather than
making the more general assumptions of Ref. \cite{Takeshi}. The
result can be generalized, in a straighforward way, following the
development of Ref. \cite{Takeshi}.  We note further than our main
result is mentioned in Ref. \cite{Takeshi}, although it is not
investigated in detail there.

The plan of our paper is as follows: In section \ref{wexp} we
construct the model of scalar field evolution near a stable
potential extremum, and we derive the expression for $w(a)$. In
section \ref{Comparison} we apply our formula to various
potentials, comparing the results to the exact evolution arising
by numerical integration. We then constrain this
general family of models with SNIa observations.
Our results are summarized in section \ref{Concl}.

\section{Field evolution near a stable potential-extremum}
\label{wexp}

We are interested in models where the scalar field evolves near a
stable potential-extremum, and thus we consider a minimally
coupled scalar field $\phi$ in a potential $V(\phi)$, where the
field $\phi$ can be either a canonical or a phantom one. In the
following we introduce the usual $\ve$-parameter, acquiring the
value $+1$ for the canonical case, and $-1$ for the phantom one.
Doing so we can present our expressions in a general way.

The Euler-Lagrange equation of motion of the field reads: \be
\label{KG}
 \ddot{\phi}+3H\dot{\phi}+\ve\frac{dV}{d\phi}=0,
 \ee where
$a$ is the scale factor and $H\equiv\dot{a}/a$ is the expansion
rate. Dots denote derivatives with respect to time and primes
denote derivatives with respect to the field $\phi$. In a flat
universe, the expansion rate is linked to the total density
$\rho_{\rm T}$ via the first Friedman equation (in units where
$8\pi G=1$) as
\be
 H^2=\rho_{\rm T}/3.
\ee Additionally, the evolution of the scale factor is given by:
\be
\frac{\ddot{a}}{a}=-\frac{1}{6}\(\rho_{\rm T}+p_{\rm T}\),
 \ee
where $p_{\rm T}$ is the total pressure.

Following
 \cite{DuttaScherrer} we perform the transformation
\be
 \label{uphi}
  \phi(t)=u(t)/a(t)^{3/2},
 \ee
 and therefore Eq.~(\ref{KG}) becomes
  \be
\label{KGnew}
 \ddot{u}+\frac{3}{4}p_{\rm
T}u+\ve a^{3/2}V'\(u/a^{3/2}\)=0. \ee

We consider a universe consisting of pressureless matter and a
scalar field, where the scalar field plays the role of the dark
energy. In order to realistically mimic the observed dark energy,
the scalar field must have $w$ close to $-1$ and its energy
density must be roughly constant. The total pressure $p_{\rm T}$
is then simply given by $p_{\rm T}\approx -\rho_{\phi0}$, where
$\rho_{\phi0}$ is the present day density of the dark energy. (In
what follows, a subscript $0$ always indicates a present day
value). Under this approximation, \eqn{KGnew} becomes: \be
\label{KGu} \ddot{u}-\frac{3}{4}\rho_{\phi 0}u+\ve
a^{3/2}V'\(u/a^{3/2}\)=0. \ee
 Since we are interested in applying \eqn{KGu} to a scalar
field evolving near a local potential-extremum  at $\phi_{*}$, for
any $\phi$ close $\phi_{*}$ we use the expansion:
 \be
V\(\phi\)=V\(\phi_{*}\)+\(1/2\)V''\(\phi_{*}\)\(\phi-\phi_*\)^2+O\(\(\phi-\phi_*\)^3\).
\ee
 Substituting the above relation into \eqn{KGu}, and imposing
$V\(\phi_{*}\)=\rho_{\phi 0}$ we obtain the following differential
equation for the field evolution:
 \be
 \label{unew}
\ddot{u}-\[\(3/4\)V\(\phi_{*}\)-\ve V''\(\phi_{*}\)\]u=0.
 \ee
This is essentially
the same equation derived previously in Ref. \cite{DuttaScherrer}.

In this work we examine cosmological evolution near a stable
potential-extremum, that is either a canonical field ($\ve>0$)
near a local minimum ($V''(\phi_*)>0$), or a phantom field
($\ve<0$) near a local maximum ($V''(\phi_*)<0$). Thus, in both
cases of interest  $\ve V''(\phi_*)>0$ and therefore the following
analysis can be performed in a unified way.

Defining
\be
\label{k}
 k\equiv\sqrt{\(3/4\)V\(\phi_{*}\)-\ve
V''\(\phi_{*}\)},
 \ee
   we obtain
the general solution of \eqn{unew} as
 \be
 \label{usol}
 u=A\sinh\(kt\)+B\cosh\(kt\),
  \ee
  where $A$ and $B$ are arbitrary constants.
  Note that  this solution holds for $k$
being either real or imaginary, where in the latter case the
hyperbolic functions are straightforwardly replaced
by trigonometric ones.  Hence, we can simply use the earlier results
of Ref. \cite{DuttaScherrer}.

Defining
\be t_{\Lambda}\equiv 2/\sqrt{3\rho_{\phi
0}}=2/\sqrt{3V\(\phi_{*}\)}, \ee
and taking $\phi\(t=0\)=\phi_{i}$, we obtain
\be
 \label{phishort}
 \phi=\frac{\phi_i}{kt_{\Lambda}}\frac{\sinh\(kt\)}{\sinh\(t/t_{\Lambda}\)},
\ee
and the equation of state parameter, $w(a)$, is given by
(see Ref. \cite{DuttaScherrer} for the details):
\begin{widetext}
\be
\label{final EOS}
1+w(a)=(1+w_0)a^{-3}
\frac{{\left[ {\sqrt {\Omega _{\phi 0}} kt_\Lambda\cosh
\left[ {k t\left( a \right)} \right] - \sqrt {(1-\Omega_{\phi 0})a^{-3}
+\Omega_{\phi 0}}
\sinh \left[ {k t\left( a \right)} \right]} \right]^2 }}
{{\left[ {\sqrt {\Omega_{\phi 0}}  kt_\Lambda\cosh ( {kt_0})
 - \sinh ( {k t_0})} \right]^2 }},
\ee
\end{widetext}
where $w_0$ is the present-day
value of $w(a)$, and $t(a)$ and $t_0$ can be derived from:
\begin{equation}
\label{ta}
t(a) = t_\Lambda \sinh^{-1}\sqrt{\left(\frac{\Omega_{\phi 0} a^3}
{1-\Omega_{\phi 0}}\right)}
\end{equation}
and
\begin{equation}
\label{t0}
t_0 = t_\Lambda \tanh^{-1} \left(\sqrt{\Omega_{\phi 0}}\right).
\end{equation}
Note that in (\ref{final EOS}) the $\ve$-parameter has been
simplified in favor of $w_0$ which is obviously smaller than $-1$
for phantom while it is larger than $-1$ for the quintessence
case.

Following Ref. \cite{DuttaScherrer}, we now
introduce the constant $K\equiv kt_{\Lambda}$. In terms of the
 potential, $K$ can be written as
\begin{equation}
\label{Kdef} K = \sqrt{1 - \ve\frac{4V^{\prime
\prime}(\phi_*)}{3V(\phi_*)}}.
\end{equation}
In terms of $K$ we can finally express the evolution of $w$ in the
following form:
\begin{widetext}
\begin{equation}
\label{finalfinal}
1 + w(a) = (1+w_0)a^{3(K-1)}\frac{[(F(a)+1)^K(K-F(a))
+(F(a)-1)^K(K+F(a))]^2}
{[(\Omega_{\phi0}^{-1/2}+1)^K(K-\Omega_{\phi0}^{-1/2})
+(\Omega_{\phi0}^{-1/2}-1)^K (K+\Omega_{\phi0}^{-1/2})]^{2}},
\end{equation}
\end{widetext}
where $F(a)$ is given by
\be
F(a) = \sqrt{1+(\Omega_{\phi 0}^{-1}-1)a^{-3}}.
\ee
(Note that $F(a) = 1/\sqrt{\Omega_\phi(a)}$, where $\Omega_\phi(a)$
is the value of $\Omega_\phi$ as a function of redshift, so that
$F(a=1) = \Omega_{\phi0}^{-1/2}$.)

Expression (\ref{finalfinal}) is identical in form to the
expression for $w(a)$ for a quintessence  field near a local
maximum  \cite{DuttaScherrer} or phantom near a local minimum
\cite{DuttaScherrer1} (and see Ref. \cite{Takeshi}
for the derivation of this expression for more general
cases).  However, the crucial difference lies in the
definition of $K$ (equation \ref{Kdef}). In the
cases considered in Refs. \cite{DuttaScherrer} and \cite{DuttaScherrer1},
$K$ was
always real since  $\ve V^{\prime \prime}(\phi_*)<0$ (in the
present language). In the case we are considering here,
for which $\ve
V^{\prime \prime}(\phi_*)>0$, $K$ can be real or imaginary (see also
\cite{Takeshi}), corresponding to
$\varepsilon V^{\prime \prime}/V < 3/4$ or $\varepsilon V^{\prime \prime}/V > 3/4$,
respectively.  Further, for the special case of
$K=0$, equations (\ref{phishort}) and (\ref{finalfinal}) are no
longer valid; instead we have
\be
 \label{phishort0}
 \phi=\frac{\phi_i}{t_{\Lambda}}\frac{t}{\sinh\(t/t_{\Lambda}\)}
\ee
and
\begin{widetext}
\be \label{final EOS K=0}
 1+w(a)=(1+w_0)a^{-3}
 \frac{{\left[ {\sqrt
{\Omega _{\phi 0}}  -  \sqrt {(1-\Omega_{\phi 0})a^{-3}
+\Omega_{\phi 0}} } \sinh^{-1}\left(\sqrt{\frac{a^3\Omega _{\phi
0}}{1-\Omega _{\phi 0}}}\right) \right]^2 }} {{\left[ {\sqrt
{\Omega_{\phi 0}}
 - \sinh^{-1}\left(\sqrt{\frac{a^3\Omega _{\phi
0}}{1-\Omega _{\phi 0}}}\right)} \right]^2 }}.
 \ee
\end{widetext}

Now consider the behavior of the scalar field for the three cases
of interest, $K^2 > 0$, $K = 0$, and $K^2 < 0$.  When $K^2 > 0$,
the functional form of equation (\ref{finalfinal}) for
quintessence evolution near a potential minimum (or phantom
evolution near a local maximum) is identical to the form for the
evolution of quintessence near a maximum (or phantom near a
minimum) given in Refs. \cite{DuttaScherrer,DuttaScherrer1}, or
for the evolution described in Ref. \cite{Takeshi}.  However, the
crucial difference is that in the previously-considered cases, we
have $K > 1$, while for the case considered here, we have instead
$0 < K < 1$ (the special case $K=1$ is discussed in Ref.
\cite{Takeshi}). Therefore, although qualitatively the behavior of
$w(a)$ is similar to that in Refs.
\cite{DuttaScherrer,DuttaScherrer1,Takeshi}, i.e. $w(a)$ decreases
monotonically, there are significant quantitative differences.

When $K=0$, we can no longer use equation (\ref{finalfinal}), but
equation (\ref{final EOS K=0}) gives an evolution for $w(a)$ that
is qualitatively similar to the $K^2 > 0$ case, i.e., a slow
monotonic decrease in $w$.

Finally, when
$K^2 < 0$, $w(a)$ is oscillatory. While equation (\ref{finalfinal})
is formally valid (and gives a real expression for $w(a)$) in this
case, the oscillatory behavior becomes
more transparent by writing $K=i t_\Lambda \gamma$, where $\gamma$
is real, and simplifying \eqn{final EOS} to a more intuitive form
(see also Ref. \cite{Takeshi}):
\begin{widetext}
\begin{equation}
 \label{final EOS2}
  1+w(a)=(1+w_0)a^{-3} \frac{{\left[ {\sqrt
{\Omega _{\phi 0}} \gamma t_\Lambda\cos \left[ {\gamma t\left( a
\right)} \right] - \sqrt {(1-\Omega_{\phi 0})a^{-3} +\Omega_{\phi
0}} \sin \left[ {\gamma t\left( a \right)} \right]} \right]^2 }}
{{\left[ {\sqrt {\Omega_{\phi 0}}  \gamma t_\Lambda\cos ( {\gamma
t_0})
 - \sin ( {\gamma t_0})} \right]^2 }}.
\end{equation}
\end{widetext}

\begin{figure}[h]
    \epsfig{file=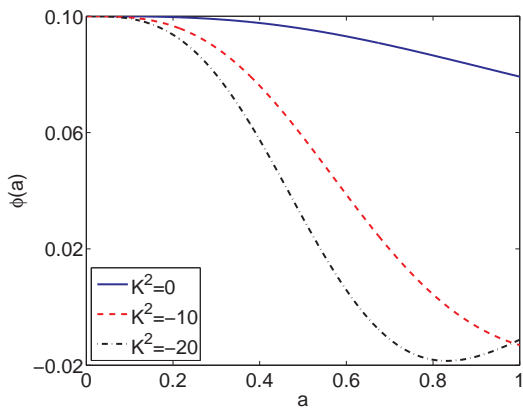,height=55mm}
    \caption
    {   \label{phievol} The evolution of the scalar field $\phi$ as a function
    of the scale factor $a$, for the indicated values of $K$, as defined
    in Eq. (\ref{Kdef}).}
\end{figure}

The behavior of $\phi(a)$ for several different values of $K$
is illustrated in Fig. (\ref{phievol}).
The quantity $K$ can be physically interpreted as a measure of the sharpness of
the curvature of the potential at its extremum. From the
definition of $K$ (Eq. \ref{Kdef}) we note that in order for
$K$ to be real, $|V''|/V$ is small, implying a ``flat'' potential.
The field evolves in a slow, friction-dominated manner (with the
Hubble parameter acting as the friction coefficient according to
Eq. \ref{KG}), asymptotically coming to rest at the potential
extremum (minimum for quintessence and maximum for phantom). On the
other hand, an imaginary $K$ requires a large $|V''|/V$, implying
a sharp curvature of the potential at the extremum, which allows
for oscillations since the friction term can be overcome.

It is important to note that we fix the minimum
value of the potential to be at  the
level of the cosmological constant, so the evolution is still
potential energy dominated (i.e., $w\simeq-1$) as the field
oscillates around the extremum. These oscillatory solutions are therefore
very different from the ones usually considered in oscillating dark
energy models \cite{sahni,Hsu_osc,masso,gu,DuttaScherrer_osc,johnson},
where the potential minimum is fixed at $V=0$, and $w$
oscillates between $\pm 1$.

\section{Comparison to exact solutions}
\label{Comparison}

In this section, we compare our analytic expression for the
evolution of $w$ to the numerically computed  exact evolution for
a few different models. In each case we have a perfect-fluid dark
matter and a dark-energy component attributed to a quintessence
($\ve=+1$) or a phantom ($\ve=-1$) field $\phi$.
For the quintessence case, we consider three different potentials
which have local minima, and we
use the corresponding
inverted potentials for the phantom case.
Our purpose is not to propose these toy models as specific possibilities
for the dark energy, but rather to
test the accuracy of our approximation
against a variety of different possibilities.

The PNGB model \cite{Frieman}, has a
potential given by
\begin{equation}
\label{pPNGB}
V(\phi) = \rho_{\phi 0}+\ve M^4 \[1-\cos\(\phi/f\)\],
\end{equation}
where $M$ and $f$ are constants. (For
recent discussions of the PNGB
model in the context of dark energy, see, e.g., Refs. \cite{Dutta,Albrecht,LinderPNGB}
and references therein).
Other models with a local potential minimum
include the Gaussian potential,
\begin{equation}
\label{pGaussian}
V(\phi) = \rho_{\phi 0}+ \ve M^4 \[1-e^{-\phi^2/\sigma^2}\],
\end{equation}
and the quadratic potential
\begin{equation}
\label{pquadratic}
V(\phi) = \rho_{\phi 0}+ \ve V_2\phi^2.
\end{equation}
where $\sigma$ and $V_2$ are constants.
We set initial conditions deep within the matter-dominated regime.
The value of the potential at the extremum, $\rho_{\phi 0}$,
is chosen to be equal to the energy of the cosmological constant.
The initial velocity of the field is taken to be zero.

As discussed above, our formalism applies to models for which
 (\ref{slow1}) is satisfied, but  (\ref{slow2}) is not.
The initial value of the field $\phi_i$ determines the accuracy of
the first slow-roll condition (\ref{slow1}) at the initial time.
To push our formalism to its limits, we choose
$\phi_i$ to a value for which
\begin{eqnarray*}
\[\frac{1}{V}\frac{dV}{d\phi}\]_{a\rightarrow0}^2&\la & O\[1\]
\end{eqnarray*}
For smaller $\phi_i$ of course, the agreement is better.

We have examined three types of curvature of the potential at the
extremum, characterized by $K$. The ``flat'' regime is represented
by $0\leq K^2<1$, where $w$ evolves in a slow monotic manner, and
$K^2=0$ represents a limiting case of this behavior.  The
``curved'' regime is represented by $-\infty<K^2<0$  in which $w$
eventually oscillates. In this regime, we have considered the
cases of $K^2=-10$ and $K^2=-20$ for each model.

\begin{figure}
    \epsfig{file=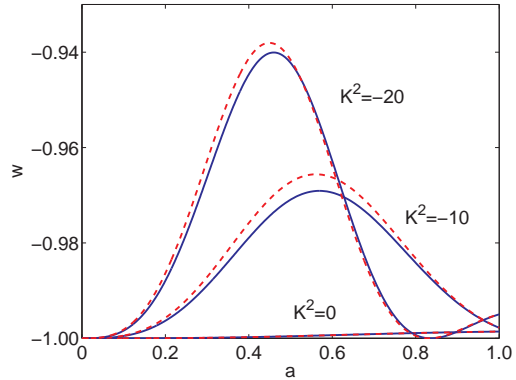,height=55mm}
    \caption
    {   \label{quadraticfig} The evolution of $w$ for quintessence in a quadratic
    potential for three different values of $K^2$. The solid blue curves indicate
     the exact evolution and the red dashed curves indicate the analytic prediction.}
\end{figure}

\begin{figure}
    \epsfig{file=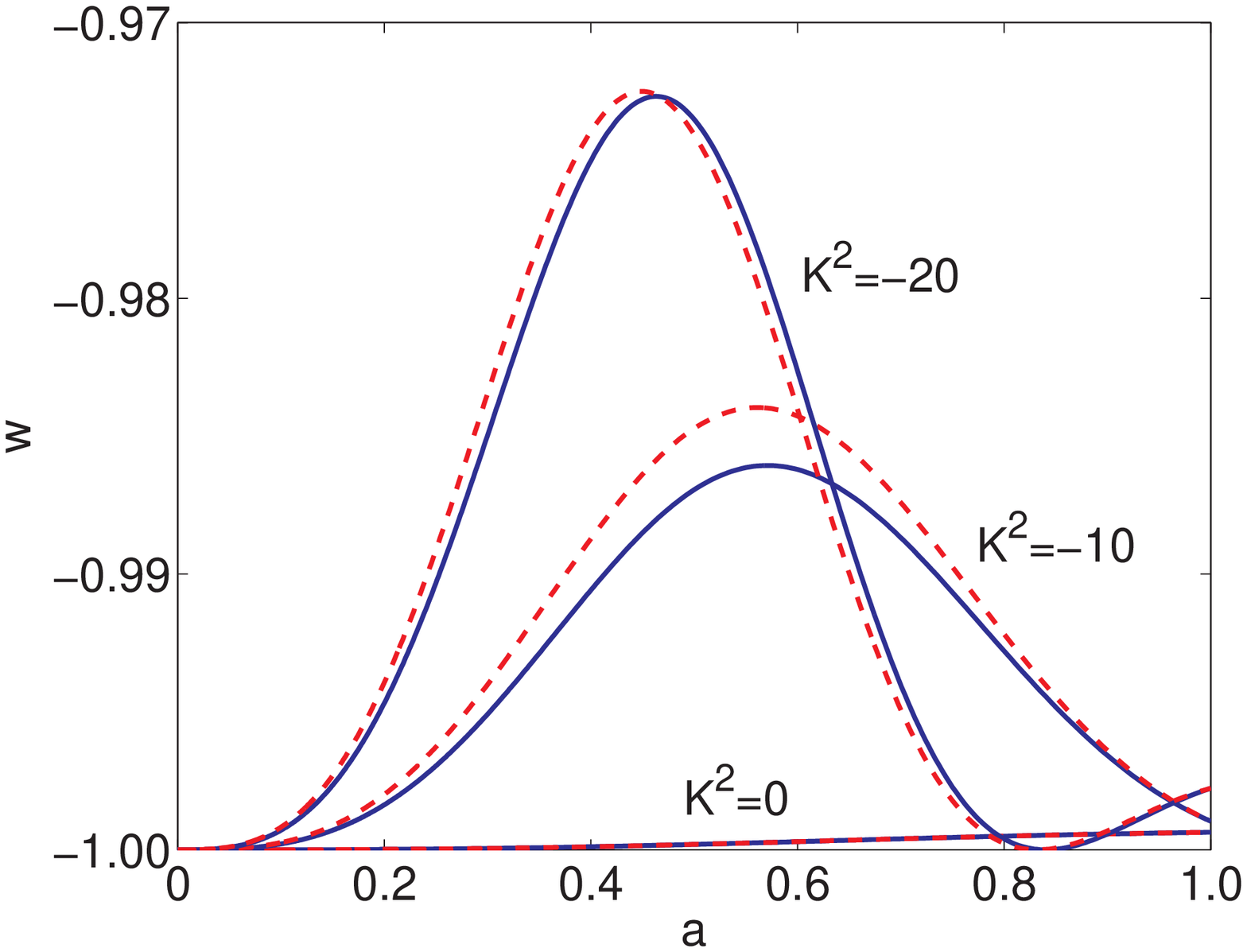,height=55mm}
    \caption
    {   \label{PNGB}  The evolution of $w$ for quintessence in a PNGB potential for three
    different values of $K^2$.  The solid blue curves indicate the exact
    evolution and the red dashed curves indicate the analytic prediction.}
\end{figure}

\begin{figure}
    \epsfig{file=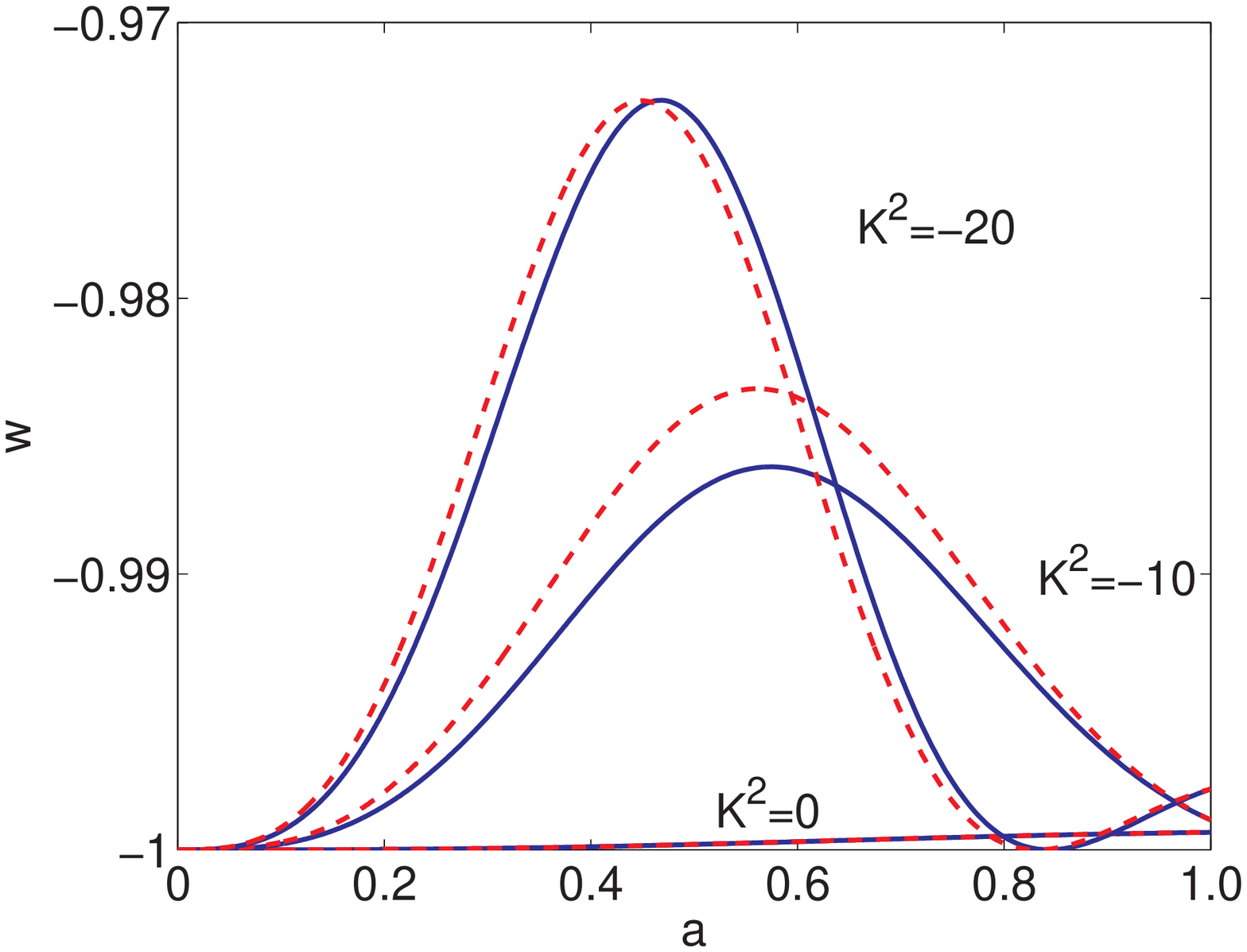,height=55mm}
    \caption
    {   \label{exponential} The evolution of $w$ for quintessence in a Gaussian
    potential for three different values of $K^2$. The solid blue curves indicate the
     exact evolution and the red dashed curves indicate the analytic prediction. }
\end{figure}

\begin{figure}
    \epsfig{file=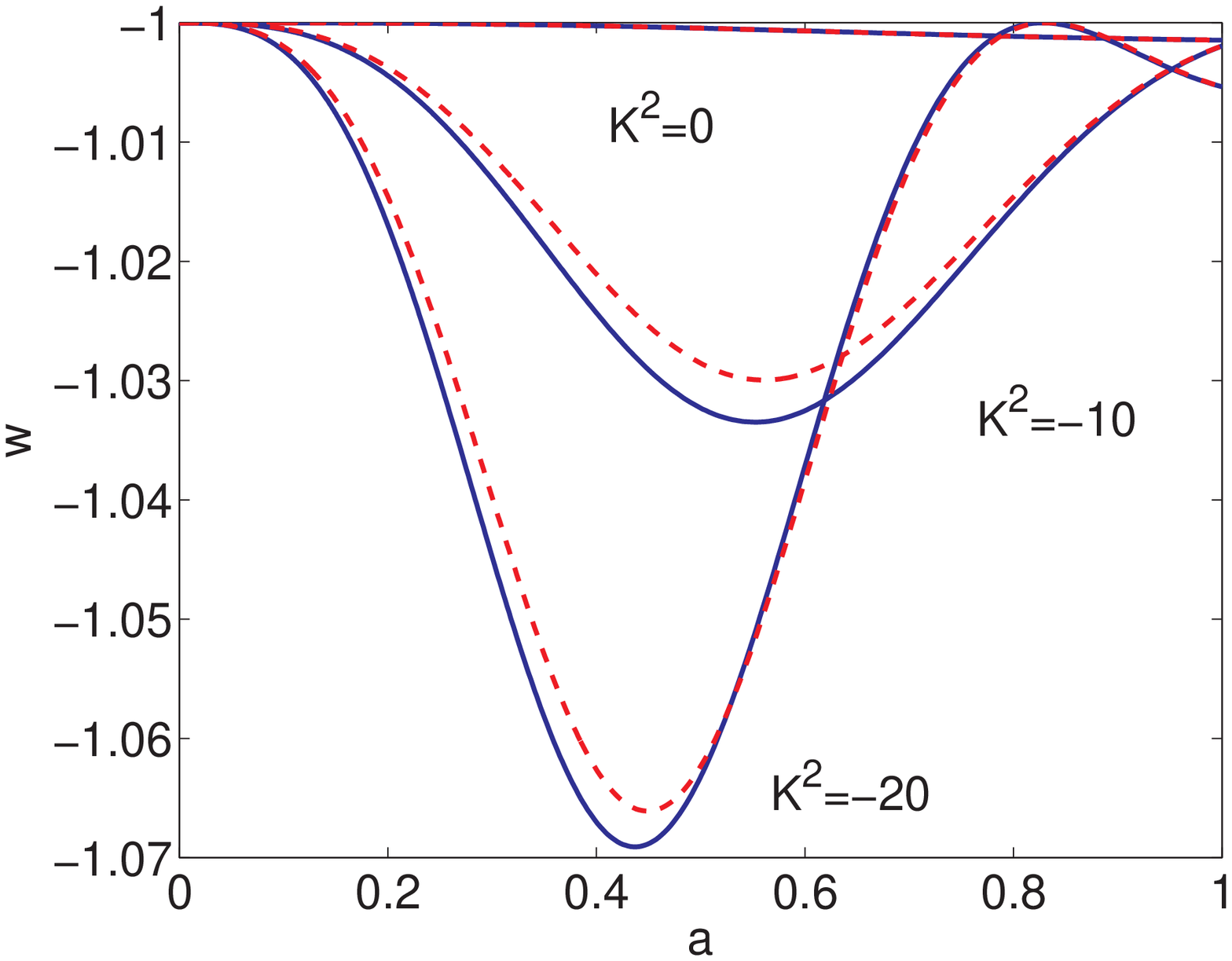,height=55mm}
    \caption
    {   \label{quadratic_p} The evolution of $w$ for a phantom in a quadratic potential
    for three different values of $K^2$. The solid blue curves indicate the exact evolution and the red dashed curves indicate the analytic prediction.}
\end{figure}

\begin{figure}
    \epsfig{file=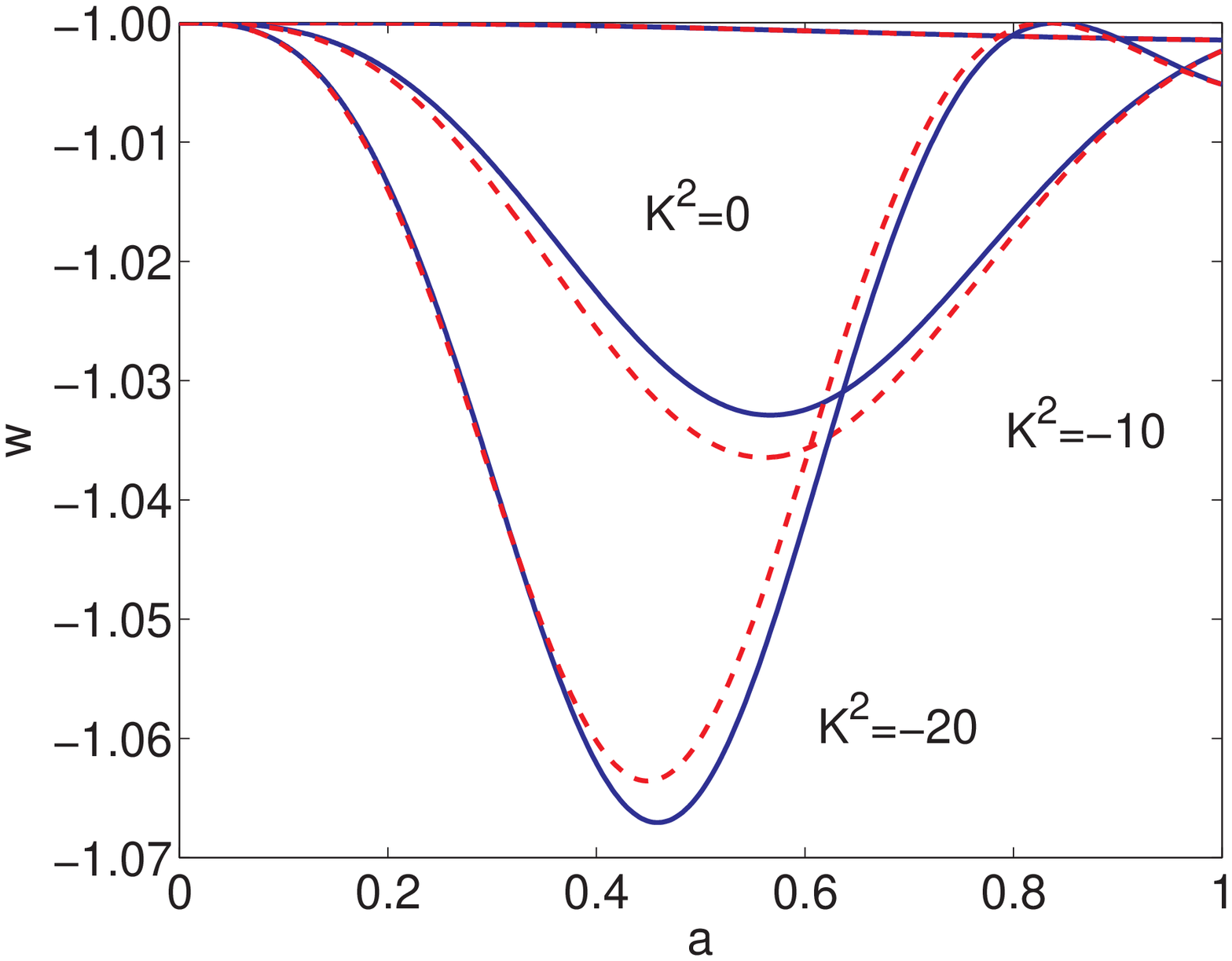,height=55mm}
    \caption
    {   \label{PNGB_p} The evolution of $w$ for a phantom in a PNGB potential
     for three different values of $K^2$.  The solid blue curves indicate the exact evolution and the red dashed curves indicate the analytic prediction.}
\end{figure}

\begin{figure}
    \epsfig{file=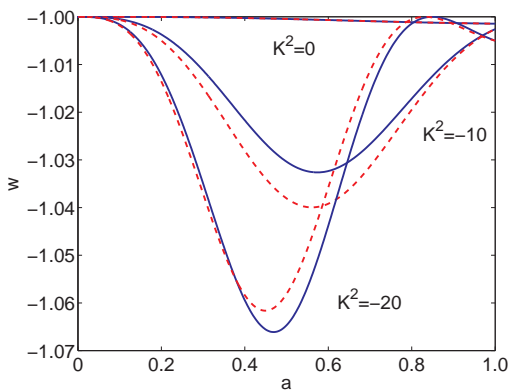,height=55mm}
    \caption
    {   \label{exponential_p} The evolution of $w$ for a phantom
    in a Gaussian potential for three different values of $K^2$.  The
    solid blue curves indicate the exact evolution and the red dashed curves indicate the analytic prediction.}
\end{figure}

In Figs. (\ref{quadraticfig}-\ref{exponential_p}), the evolution
of $w$ from (\ref{final EOS K=0}) (for $K^2 =0$) or (\ref{final EOS2})
(for $K^2 < 0$)
is shown in comparison to the exact
evolution for the three different models, and the different $K-$
regimes described above.  The agreement, in all
 three cases, between our analytic approximation and
the exact numerical evolution is excellent, with errors $\delta w
\alt 0.01$ in all cases.  This result indicates that our
expressions for $w(a)$ for our three cases of interest can be
considered ``generic" expressions that apply to a wide range of
possible quintessence and phantom models.

\begin{figure}
    \epsfig{file=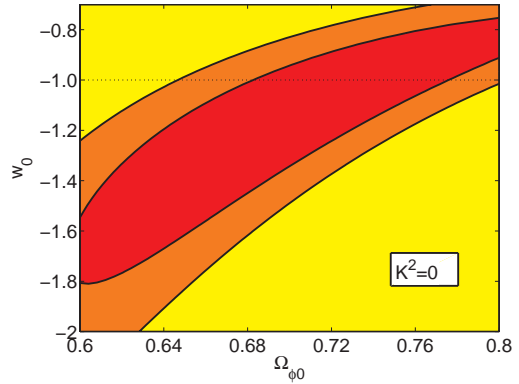,height=55mm}
    \caption
    {   \label{chis_0}Likelihood plot from SNIa data for the parameters $w_0$ and $\Omega_{\phi0}$,
for quintessence and phantom models with generic behavior
described by Eq. \eqref{final EOS K=0}, with $K^2=0$,
where $K$
is the function of the curvature of the potential
at its extremum given in Eq. \eqref{Kdef}.
The yellow (light) region
is excluded at the 2$\sigma$ level, and the darker (orange) region
is excluded at the 1$\sigma$ level.  Red (darkest) region is
not excluded at either confidence level.}
\end{figure}

\begin{figure}
    \epsfig{file=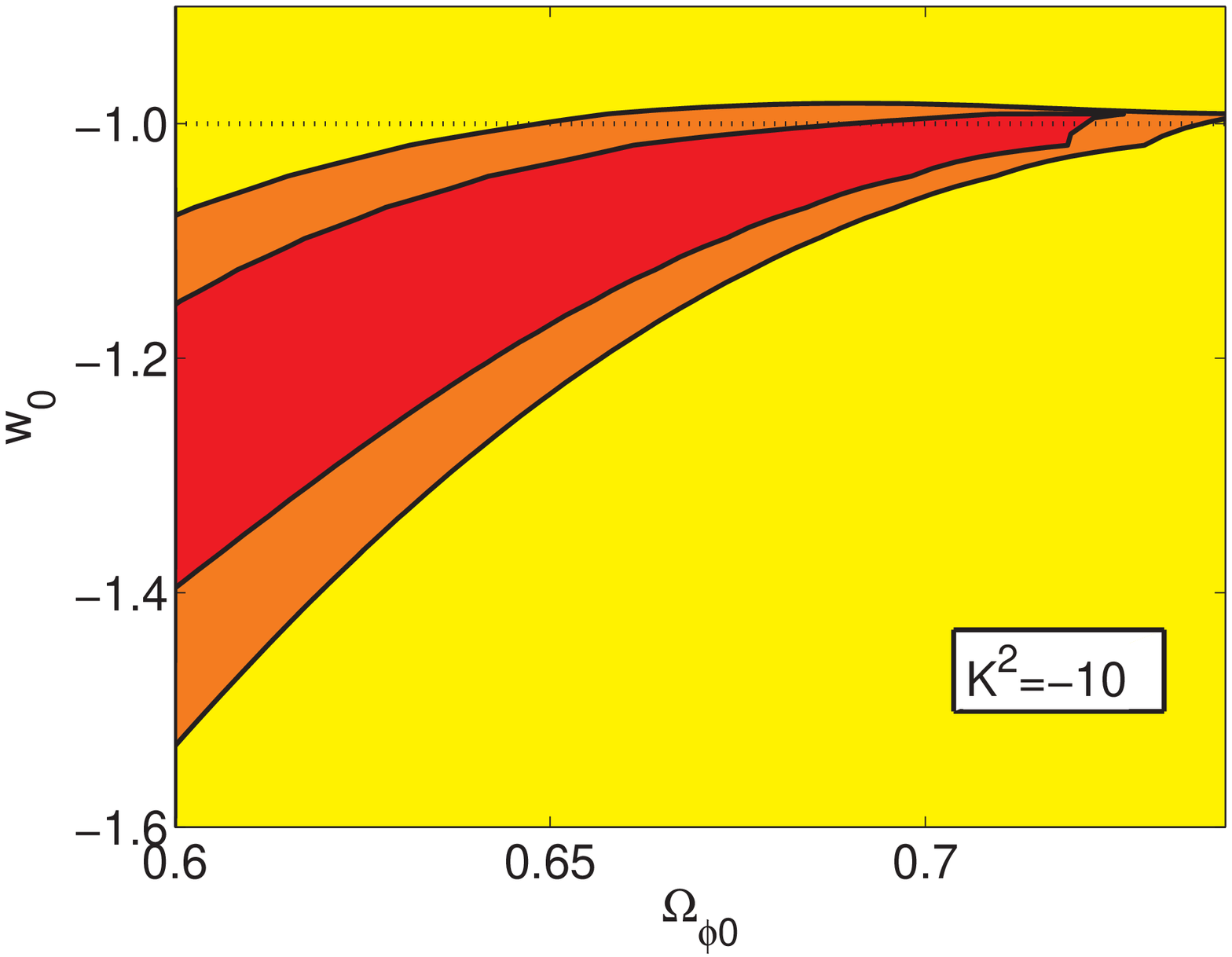,height=55mm}
    \caption
    {   \label{chis_10}Likelihood plot from SNIa data for the parameters $w_0$ and $\Omega_{\phi0}$,
for quintessence and phantom models with generic behavior
described by Eq. \eqref{final EOS2}, with $K^2=-10$,
where $K$
is the function of the curvature of the potential
at its extremum given in Eq. \eqref{Kdef}.
The yellow (light) region
is excluded at the 2$\sigma$ level, and the darker (orange) region
is excluded at the 1$\sigma$ level.  Red (darkest) region is
not excluded at either confidence level.}
\end{figure}

\begin{figure}
    \epsfig{file=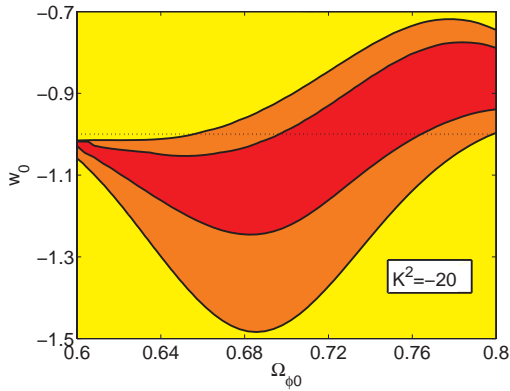,height=55mm}
    \caption
    {   \label{chis_20}Likelihood plot from SNIa data for the parameters $w_0$ and $\Omega_{\phi0}$,
for quintessence and phantom models with generic behavior
described by Eq. \eqref{final EOS2}, with $K^2=-20$,
where $K$
is the function of the curvature of the potential
at its extremum given in Eq. \eqref{Kdef}.
The yellow (light) region
is excluded at the 2$\sigma$ level, and the darker (orange) region
is excluded at the 1$\sigma$ level.  Red (darkest) region is
not excluded at either confidence level.}
\end{figure}

Finally, in Figs.~(\ref{chis_0}-\ref{chis_20}), we use our
analytic approximation (Eqs. \ref{final EOS K=0}-\ref{final EOS2}) to construct a $\chi^2$
likelihood plot for $w_0$ and $\Omega_{\phi0}$ for the three
choices of $K^2$ using the recent Type Ia Supernovae standard
candle data (ESSENCE+SNLS+HST from \cite{Davis}). We have
exploited the fact that our expressions for $w(a)$ for
quintessence and phantom models have the same functional form,
allowing us to produce a likelihood plot that is continuous across
$w_0 = -1$.  However, it is important to note that in these
figures, the dashed line at $w_0 = -1$ divides two distinct
models. Clearly, these models are not ruled out by current
supernova data.

\section{Conclusions}
\label{Concl}

Using techniques previously applied in
\cite{DuttaScherrer,DuttaScherrer1,Takeshi}, we have derived a general
expression for the evolution of $w$, which is valid for a wide
class of quintessence and phantom dark energy models in which the
field is rolling close to a stable local potential extremum (i.e.,
a minimum for the quintessence and a maximum for the phantom case,
respectively). Such models provide a mechanism to produce a value
of $w$ that is close to $-1$. We have
tested our expression against the (numerically determined) exact
evolution for three different representative models and in each case it
replicated the exact evolution studied with an accuracy greater
than $99\%$. A comparison between our generic approximation and
the observational data indicates that these models are allowed by
SNIa data for a variety of values of the potential curvature parameter $K$
defined in equation (\ref{Kdef}).

Finally, we note that the case considered here, in which $V^{\prime\prime} > 0$ for quintessence
(and $V^{\prime\prime}<0$ for phantom models) leads to a much richer set
of behaviors than the previously-examined case of $V^{\prime\prime} <0$
for quintessence (and $V^{\prime\prime} > 0$ for phantom models).  In the
case examined here, we see three very different regimes, depending
on the sign of $K^2$.

\bigskip
\bigskip
\bigskip

\acknowledgments We thank T. Chiba for helpful discussions.
S.D. and R.J.S. was supported in part by the Department of
Energy (DE-FG05-85ER40226).

\end{document}